\documentclass[10pt,prl,aps,twocolumn,superscriptaddress]{revtex4-1}

\pdfoutput=1
\usepackage{graphicx}
\usepackage{amssymb}
\usepackage{times}
\usepackage{amsmath}
\usepackage{amsfonts}
\usepackage{txfonts}
\usepackage{color}
\usepackage{natbib}
\usepackage{units}
\usepackage{textcomp,calrsfs}
\usepackage{pdfpages}
\usepackage{ifsym}
\usepackage{bm}
\usepackage{relsize}
\usepackage[stable]{footmisc}

\newcommand{\fig}[1]{Fig.~\ref{#1}}

\begin{document}
\title{Microphotonic Forces From Superfluid Flow}

\author{D. L. McAuslan} \thanks{These authors contributed equally to this work.} \affiliation{Centre for Engineered Quantum Systems, School of Mathematics and Physics, University of Queensland, Brisbane, QLD 4072, Australia.}
\author{G. I. Harris} \thanks{These authors contributed equally to this work.} \affiliation{Centre for Engineered Quantum Systems, School of Mathematics and Physics, University of Queensland, Brisbane, QLD 4072, Australia.} 
\author{C. Baker} \affiliation{Centre for Engineered Quantum Systems, School of Mathematics and Physics, University of Queensland, Brisbane, QLD 4072, Australia.} 
\author{Y. Sachkou} \affiliation{Centre for Engineered Quantum Systems, School of Mathematics and Physics, University of Queensland, Brisbane, QLD 4072, Australia.}
\author{X. He} \affiliation{Centre for Engineered Quantum Systems, School of Mathematics and Physics, University of Queensland, Brisbane, QLD 4072, Australia.}
\author{E. Sheridan} \affiliation{Centre for Engineered Quantum Systems, School of Mathematics and Physics, University of Queensland, Brisbane, QLD 4072, Australia.}
\author{W. P. Bowen} \affiliation{Centre for Engineered Quantum Systems, School of Mathematics and Physics, University of Queensland, Brisbane, QLD 4072, Australia.}

%%%%%%%%%%%%%%%%%%%%%%%%%%%%%%%%%%%%%%%%%%%%%%%%%%%%%%%%%%%%%%%%%%%%%%%%%%%%%%%%%%%%%%%%%%%%%%%%%%%%%%%%
%ABSTRACT
%%%%%%%%%%%%%%%%%%%%%%%%%%%%%%%%%%%%%%%%%%%%%%%%%%%%%%%%%%%%%%%%%%%%%%%%%%%%%%%%%%%%%%%%%%%%%%%%%%%%%%%%
\begin{abstract}
In cavity optomechanics, radiation pressure and photothermal forces are widely utilized to cool and control micromechanical motion, with applications ranging from precision sensing and quantum information to fundamental science. Here, we realize an alternative approach to optical forcing based on superfluid flow and evaporation in response to optical heating. We demonstrate optical forcing of the motion of a cryogenic microtoroidal resonator at a level of 1.46~nN, roughly one order of magnitude larger than the radiation pressure force. We use this force to feedback cool the motion of a microtoroid mechanical mode to 137~mK. The photoconvective forces demonstrated here provide a new tool for high bandwidth control of mechanical motion in cryogenic conditions, and have the potential to allow efficient transfer of electromagnetic energy to motional kinetic energy. 
\end{abstract}
\maketitle

%%%%%%%%%%%%%%%%%%%%%%%%%%%%%%%%%%%%%%%%%%%%%%%%%%%%%%%%%%%%%%%%%%%%%%%%%%%%%%%%%%%%%%%%%%%%%%%%%%%%%%%%%%
%INTRO
%%%%%%%%%%%%%%%%%%%%%%%%%%%%%%%%%%%%%%%%%%%%%%%%%%%%%%%%%%%%%%%%%%%%%%%%%%%%%%%%%%%%%%%%%%%%%%%%%%%%%%%%%%
%The exploitation of optical forces in microscopic systems has remained an active field of research for many decades, from the fundamental perspective of light matter interactions~\cite{Aspelmeyer_14RMP} to the practical advances enabled by mass/force/spin sensors~\cite{Metcalfe_APR14}. 
%%
%In this context a variety of optically mediated forces have been studied in optomechanical systems, 

Optical forces are widely utilized in photonic circuits~\cite{Li_Nat08, Roels_NatNano09}, micromanipulation~\cite{Ashkin_Science87, MacDonald_Nat03}, and biophysics~\cite{Burg_Nat07, Taylor_NatPhot13}. In cavity optomechanics, in particular, optical forces enable cooling and control of microscale mechanical oscillators that can be used for ultrasensitive detection of forces, fields and mass~\cite{Mamin_APL01, Forstner_PRL12, Chaste_NatNano12}, quantum and classical information systems~\cite{Beugnon_NatPhys07}, and fundamental science~\cite{Orzel_Science01, Greiner_Nature02}. Recent progress has seen radiation pressure used for coherent state-swapping~\cite{Verhagen12_Nat}, ponderomotive squeezing~\cite{Brooks12_Nat} and ground state cooling~\cite{Chan11_Nat}, while static gradient forces have enabled all-optical routing~\cite{Rosenberg_NatPhot09} and non-volatile mechanical memories~\cite{Bagheri11_NatNano}. Likewise, photothermal forces, where the mechanical element moves in response to mechanical stress from localized optical absorption and heating, have been used to demonstrate cavity cooling of a semiconductor membrane~\cite{Usami_Nat12, Barton_NanoLett12}, single molecule force spectroscopy~\cite{Stahl_RSI09} and rich chaotic dynamics in suspended mirrors~\cite{Marino_PRE11}. 
%
%More generally, radiation pressure has also enabled rapid progress in fields outside the scope of optomechanics, for example optical tweezers\cite{Ashkin_PRL70} and trapped atoms\cite{Chu_PRL86}.
%% 
%These optically mediated forces are typically extremely inefficient at converting photon energy to mechanical energy, with state of the art photothermal based AFM systems operating with efficiencies of $\eta \approx 10^{-5}-10^{-4}$ \cite{Kiracofe_RSI11}.
%Mention optical tweezers and cold atoms to broaden base and link to first paragraph

\begin{figure}[ht!]
\begin{center}
\includegraphics[width=0.95\columnwidth]{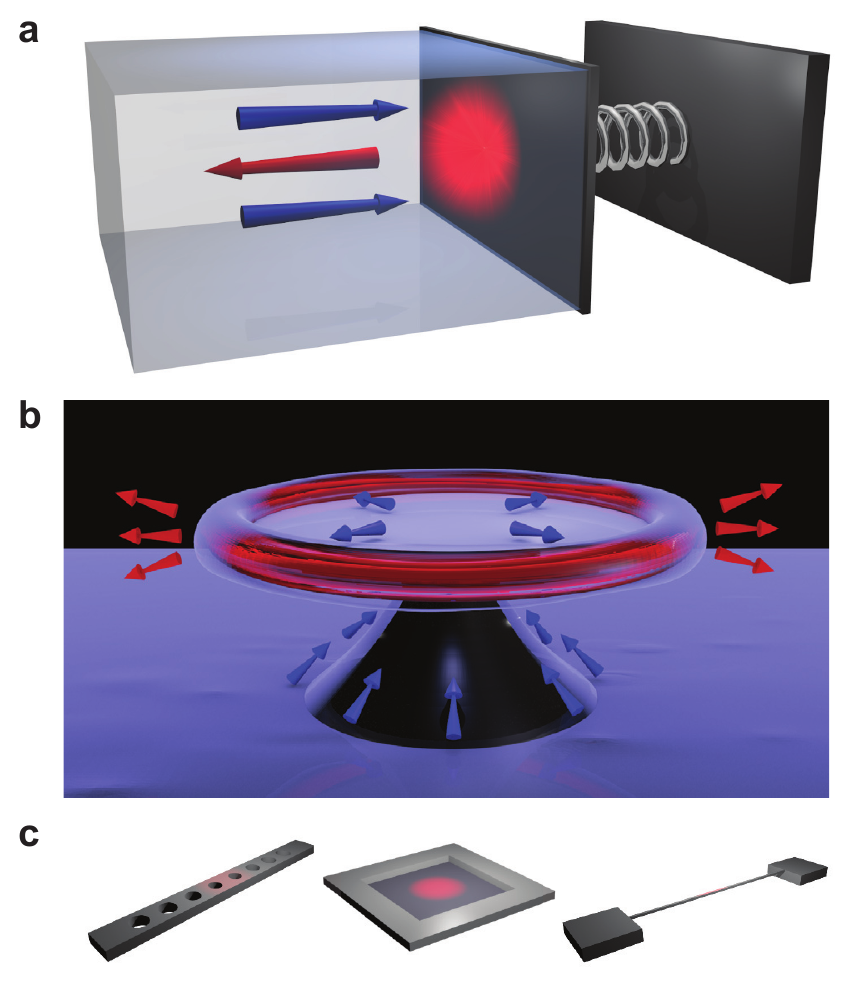}
\end{center}
\caption{\label{fig4} (a) In bulk, a local heat source generates flow of superfluid helium (blue arrows) and counter-flow of normal fluid (red arrow), imparting momentum onto the oscillator. (b) Representation of a microtoroid covered in a thin film of superfluid helium. Heat around the periphery caused by optical absorption generates fluid flow (blue arrows). At low pressures the superfluid then transitions directly into gas phase and leaves the subsystem (red arrows). (c) Superfluid mediated photothermal forcing may be readily extended to other optomechanical systems such as photonic crystal cavities, membranes and nanostrings.}
\end{figure}

Here we demonstrate an alternative photoconvective approach to optical forcing. In our implementation, this technique utilizes the convection in superfluids, whereby frictionless fluid flow is generated in response to a local heat source. This well-known superfluid fountain effect~\cite{Allen_Nat38} is a direct manifestation of the phenomenological two-fluid model proposed by Landau and Tisza~\cite{Landau_USSR41, Tisza_Nat38}. The momentum carried by the helium-4 flow is then transferred to a mechanical element via collision and recoil of superfluid atoms.
If the heat source is localized upon the mechanical element, the incident superfluid atoms are converted either to a normal fluid counter-flow or evaporated (see Fig.~\ref{fig4}(a,~b)). Alternatively, a distant heat source could be utilized with the mechanical element acting to reroute the superflow. 
%A representation of this mechanism in bulk superfluid is shown in \fig{fig4}(a). 

%In addition, the ability to spatially separate the heat source from the mechanical element gives an additional degree of freedom which is unavailible in other photothermal systems. 

%%%%%%%%%%%%%%%%%%%%%%%%%%%%%%%%%%%%%%%%%%%%%%%%%%%%%%%%%%%%%%%%%%%%%%%%%%%%%%%%%%%%%%%%%%%%%%%%%%%%%%%%%%
%Evaporative force
%%%%%%%%%%%%%%%%%%%%%%%%%%%%%%%%%%%%%%%%%%%%%%%%%%%%%%%%%%%%%%%%%%%%%%%%%%%%%%%%%%%%%%%%%%%%%%%%%%%%%%%%%%

\begin{figure}[t]
\begin{center}
\includegraphics[width=0.95\columnwidth]{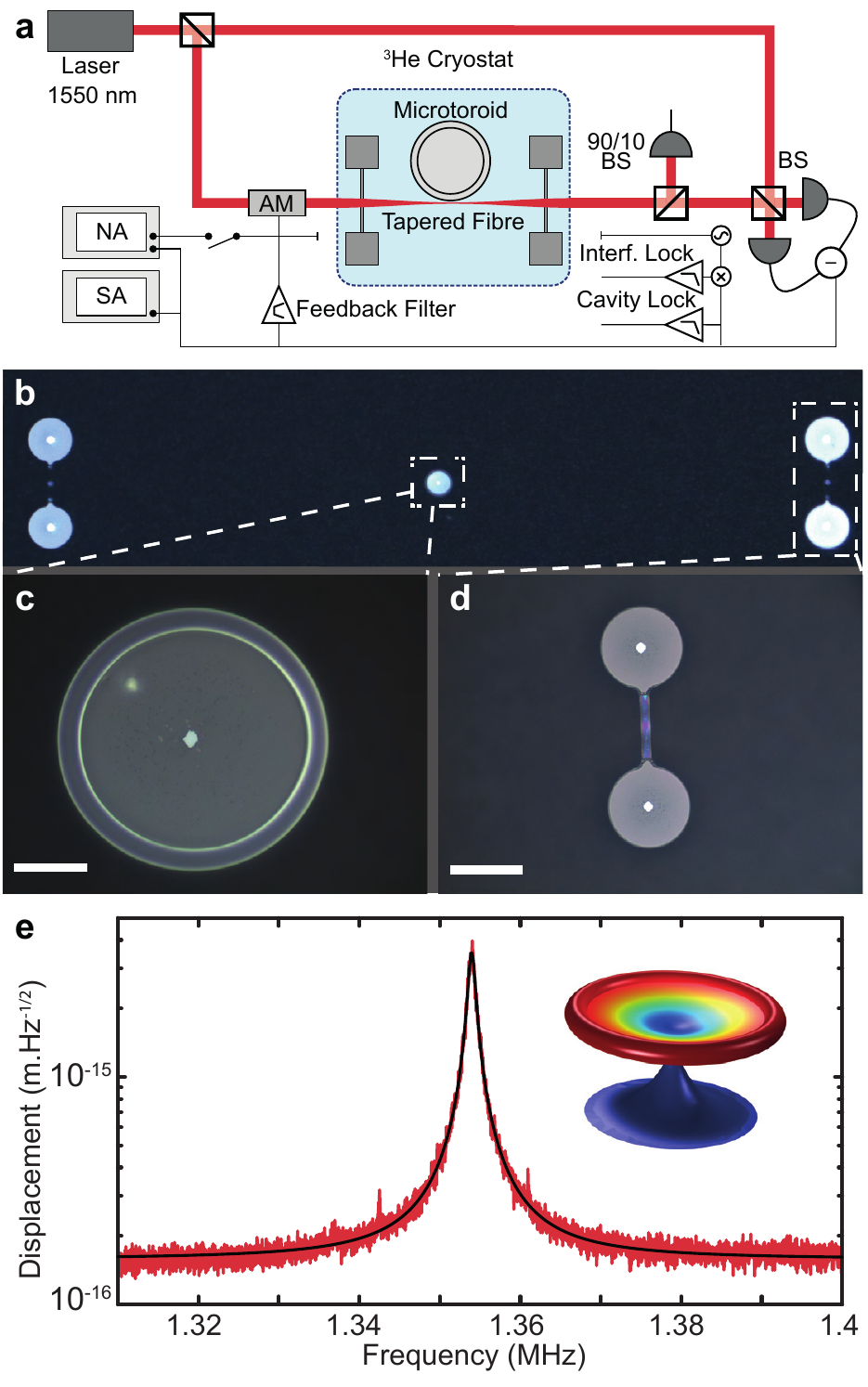}
\end{center}
\caption{\label{Fig1} (a) Experimental schematic. A microtoroid is nested inside an all-fiber interferometer and cooled by a He-3 refrigerator. BS: Beamsplitter, AM: Amplitude modulator, SA: Spectrum analyser, NA: Network analyser. (b) Optical microscope image of the microtoroid used in these experiments, showing the support beams, one either side, that are used to stabilize the tapered optical fiber. (c) Zoomed in microscope image of the microtoroid. Scale bar is $20~\mu$m long. (d) Zoomed in microscope image of a stabilization beam. The circular pads support a suspended beam which has been thinned to 200~nm thickness in order to minimize optical scattering loss. Scale bar is $100~\mu$m long. (e) Thermal motion of the flexural mode of a microtoroid at 3~K. Inset: FEM simulation of the mechanical displacement profile.}
\end{figure}

The configuration used here to realize superfluid photoconvective forcing is represented in \fig{fig4}(b). A microtoroidal resonator is covered in a several nanometre thick film of superfluid helium~\cite{Harris15_arxiv} which forms naturally due to van der Waals forces. Absorption of the circulating laser field at the microtoroid periphery (red glow) causes localized heating. This increase in temperature generates superfluid helium flow up the pedestal towards the heat source via the fountain effect (blue arrows). At the periphery superfluid helium is evaporated (red arrows) resulting in a force on the microtoroid that, on average, is directed radially inwards. 
%The toroid therefore experiences a net inward radial force due to the departing helium atoms. 
The magnitude of this radial force is given by
\begin{eqnarray}
F_{\mathrm{radial}} &=& -\frac{\mathrm{d} (mv_{\mathrm{radial}})}{\mathrm{d}t} \\
% &=& \dot{m} v_{\mathrm{rms}} \frac{1}{\pi^2}\int_{-\pi/2}^{\pi/2}\int_{-\pi/2}^{\pi/2}\cos(\theta)\cos(\phi)\mathrm{d}\theta\mathrm{d}\phi \\
&=& \frac{4}{\pi^2} \dot{m} v_{\mathrm{rms}} 
\label{Eqradialforcephotothermal}
\end{eqnarray}
where $\dot{m}$ is the mass flow rate of evaporated helium. The net radial velocity $v_{\rm radial}$ is calculated by integrating the contribution from isotropic evaporation in the outwards facing half-space with a root-mean-square (RMS) velocity of
\begin{equation}
v_{\mathrm{rms}}=\sqrt{\frac{3 k_{\rm B} T_{\rm evap}}{m_{\mathrm{He}}}}
\label{v_rms}
\end{equation} 
where $T_{\rm evap}$ is the temperature of evaporated atoms and $m_{\rm He}$ is the mass of a helium atom (see Supplementary Information). %In this situation there is no net force on the toroid in the vertical direction since the vaporization is isotropic. 

%To estimate the magnitude of force from superfluid evaporation we must estimate the mass flow rate of the superfluid. 
In steady state, the mass flow rate of the superfluid is determined by balancing the optical heat load with the energy dissipated through normal fluid counter-flow or evaporation of the film (See Supplementary Information for further discussion). While in bulk superfluid systems the energy dissipation is typically dominated by counter-flow, for thin films the normal fluid fraction is viscously clamped to the surface~\cite{Atkins_PR59}, and evaporation dominates. To prevent the continuous accumulation of fluid at the heat source the rate of evaporation must equal the in-flux from superfluid flow. For an absorbed optical power $P_\text{abs}$ the superfluid mass flow rate is then $\dot{m}=P_\text{abs}/(L-\langle \mu_\text{VDW} \rangle)$ where $L$ is the latent heat of vaporization and $\langle \mu_\text{VDW} \rangle$ is the van der Waals potential of the superfluid film (see Supplementary Information) and the resulting inward radial force from helium evaporation is
\begin{equation}
F_{\mathrm{radial}} = \frac{4}{\pi^2} \sqrt{\frac{3 k_{\rm B} T_{\rm evap}}{m_{\mathrm{He}}}} \frac{P_\text{abs}}{L-\langle \mu_\text{VDW} \rangle}.
\label{evapForce}
\end{equation}
Note that, similar to photothermal forces~\cite{Metzger_Nat04}, this expression is independent of the cavity finesse allowing photoconvective forces to be applied effectively where only a weak cavity, or no cavity, is present.
%Considering a specific scenario where $P_\text{abs} = 1~\rm\mu W$, the evaporative force is estimated from \eqn{evapForce} to be $F_{\rm radial} \approx 2\times 10^{-9}~\rm N$ for atoms with a 1~K evaporation temperature.
%
%Considering the situation of $P_\text{abs} = 1~\rm\mu W$ of absorbed optical power the mass flow rate is calculated to be $\dot{m}=P_\text{abs}/(L-\langle \mu_\text{VDW} \rangle) = 5\times 10^{-11}~\rm kg~s^{-1}$ where $L$ is the latent heat of vaporization and $\langle \mu_\text{VDW} \rangle$ is the van der Waals potential of the superfluid film (See Supplementary Information).  The resulting inward force from helium evaporation is then estimated from \eqn{Eqradialforcephotothermal} to be $F_{\rm radial} \approx 2\times 10^{-9}~\rm N$ for atoms with a 1~K evaporation temperature. 
In way of comparison, if the incident light is fully absorbed, the radiation pressure force is given by $F_\text{RP} = P_\text{abs}\mathcal{F}/c$ where $\mathcal{F}$ is the cavity finesse and $c$ is the speed of light.
For a 1~K superfluid evaporation temperature, the ratio $F_\text{radial}/F_\text{RP} \sim 4 \times 10^5/\mathcal{F}$, indicating that, in our configuration, the superfluid photoconvective force is similar in magnitude to the radiation pressure force from a cavity with a finesse of around 400,000. For our experimental conditions, with a finesse of $\mathcal{F}=53,000$, the superfluid force is predicted to be approximately one order of magnitude larger than radiation pressure.

To experimentally realize this prediction we use the setup shown in Fig.~\ref{Fig1}(a). A microtoroidal whispering-gallery-mode resonator (major radius = 37.5~$\mu$m, minor radius = 3.5~$\mu$m, Fig.~\ref{Fig1}(c)) is located inside the sample chamber of a helium-3 closed-cycle cryostat (Oxford Triton). Laser light at 1555.08~nm is coupled into a high-quality optical mode (linewidth $\kappa/2\pi = 23.5$~MHz) of the microtoroid via a tapered optical fiber. The tapered fiber rests on suspended stabilization beams fabricated near the microtoroid (Fig.~\ref{Fig1}(b,~d)), ensuring that cryostat vibrations do not affect the taper-toroid separation. The microtoroid supports a number of intrinsic mechanical modes ranging in frequency from 1~MHz to 50~MHz. The thermal motion of these modes is imprinted as phase fluctuations onto the optical field which are measured using homodyne detection. 
The radial forces applied by both radiation pressure and superfluid flow have optimal overlap with the radial breathing mode of the toroid, at 40~MHz. However, the superfluid forcing was observed to be ineffective above frequencies of approximately 2 MHz, possibly due to breakdown of superfluidity as the Landau critical velocity is reached~\cite{Landau_USSR41}. 
%We attribute this behaviour to the Landau critical velocity, beyond which superfluidity breaks down~\cite{Landau_USSR41}. 
%The maximum frequency at which the superfluid can respond is limited by the critical velocity and the distance it needs to travel, which in this experiment is on the order of the microtoroid radius. 
%Because the superfluid speed is limited to the critical velocity the maximum frequency at which the superfluid can respond is limited by this and the distance it needs to travel, which in this experiment is on the order of the microtoroid radius. 
Consequently, we perform experiments with the first order flexural mode at $\Omega_{\rm m}/2\pi = 1.35~\rm MHz$, which has a mechanical dissipation rate of $\Gamma_\text{m}/2\pi = 530$~Hz at base temperature (559~mK). The single-photon optomechanical coupling rate of this mode is measured as $g_0/2\pi = 12.3$~Hz and the Brownian fluctuations at $3~\rm K$ are shown in Fig.~\ref{Fig1}(e) with the displacement profile from finite element modelling (FEM) shown in the inset.
%
%The particular mechanical mode studied here is the first-order flexural mode of the microtoroid at $\Omega_{\rm m}/2\pi = 1.35~\rm MHz$, which has a mechanical dissipation rate of $\Gamma_\text{m}/2\pi = 530$~Hz at base temperature. This mode was chosen as it was observed to exhibit the strongest forcing from helium evaporation. The single-photon optomechanical coupling rate of this mode is measured as $g_0/2\pi = 12.3$~Hz and the Brownian fluctuations at $3~\rm K$ are shown in Fig.~\ref{Fig1}(b) with the displacement profile from finite element modelling (FEM) shown in the inset. 
%It was observed that the magnitude of the atomic recoil force decays at frequencies beyond 2~MHz \emph{I guess technically we havent observed this, we have observed SF forcing on the 1.35~MHz mode and much reduced forcing on a 9~MHz mode. However if you calculate the time it takes SF to move from the center of the toroid to the rim if its travelling at 70m/s, we get a rate of 2MHz}, where flow rates beyond Landau's critical velocity~\cite{Landau_USSR41} are required, and therefore superfluidity breaks down. This precludes the application of atomic recoil forces on higher frequency modes such as the radial breathing mode.

\begin{figure}[t]
\begin{center}
\includegraphics[width=0.95\columnwidth]{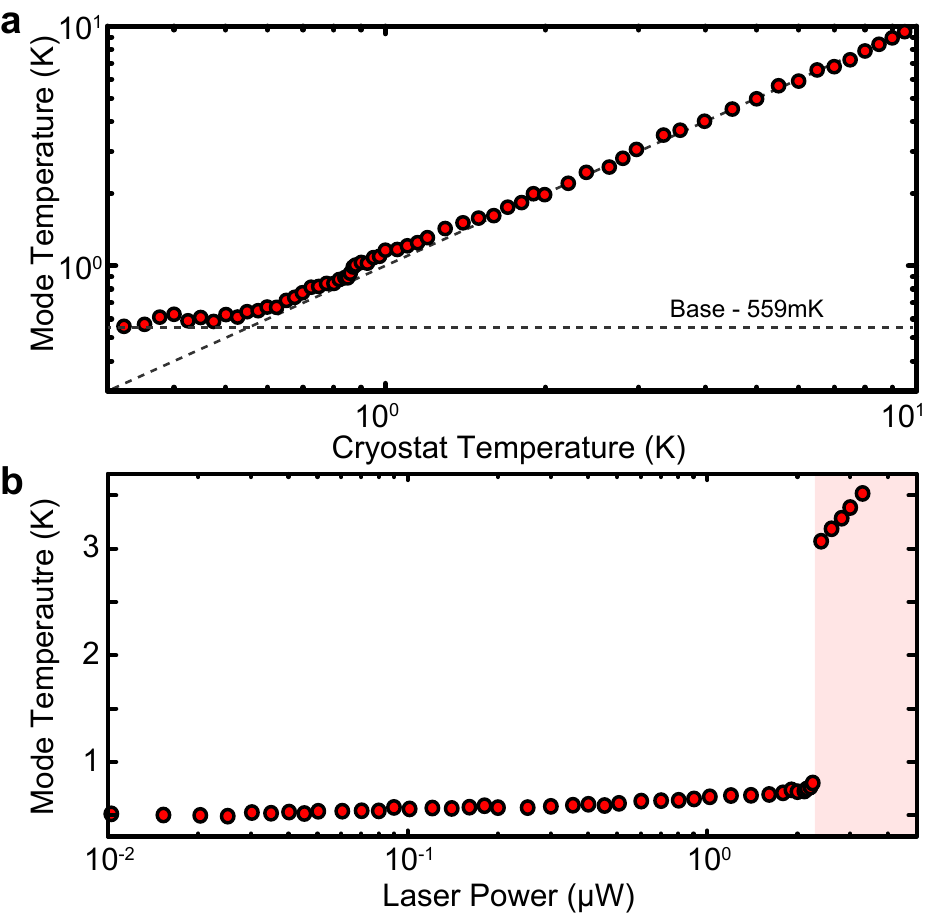}
\end{center}
\caption{\label{Fig2} %(a) Driven response of the flexural mode as the cryostat is cooled, showing a step increase in response at the superfluid transition temperature. 
(a) Mode temperature of the microtoroid flexural mode as the cryostat is cooled from 10~K to 0.32~K. The microtoroid reaches a base temperature of 0.56~K with 100~nW of injected optical power.
(b) Mode temperature of the flexural mode as the probe laser is increased from $10~\rm nW$ to $3.3~\rm \mu W$. Below $2.2~\rm \mu W$ the temperature increases slightly as the laser power is increased. Above $2.2~\rm \mu W$ the superfluid boils off causing a sharp rise in mode temperature.}
\end{figure}

To produce the superfluid film, the sample chamber was filled with low-pressure helium-4 gas (19~mBar at 2.9~K) and cooled to base temperature. This gas pressure was specifically chosen to provide a superfluid film with a thickness such that the characteristic frequencies of third sound modes intrinsic to the superfluid film~\cite{Harris15_arxiv} do not overlap with the microtoroid mode. At 850~mK the helium transitions directly from the gas phase to its superfluid state, forming a thin ($<$5~nm) superfluid layer over the chamber and its contents. To estimate the final temperature of the microtoroid the flexural mode was monitored as the cryostat temperature was decreased from 10~K to 320~mK. Spectral analysis of the homodyne photocurrent gave the mechanical mode temperature via the integrated power spectral density. From 10~K to $600~\rm mK$ the microtoroid is well thermalized to the cryostat, as shown by the linear fit in Fig.~\ref{Fig2}(a); however, at lower cryostat temperatures the microtoroid mode temperature plateaus and is no longer in thermal equilibrium with the cryostat. We attribute this temperature deviation to the heat dissipated at the sample causing a thermal gradient between the microtoroid and cryostat cold plate.

\begin{figure}[t]
\begin{center}
\includegraphics[width=0.95\columnwidth]{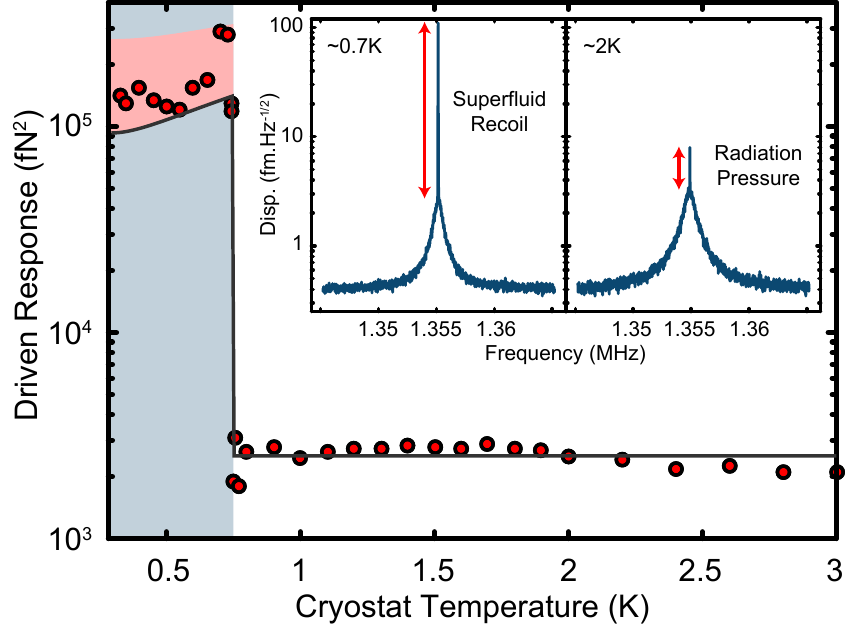}
\end{center}
\caption{\label{Fig3} Driven response of the flexural mode as the cryostat is cooled (red points), showing a step increase in response at the superfluid transition temperature. Black line represents a theoretical fit to the data. Grey shaded area indicates a superfluid layer has formed on the microtoroid surface. Pink shaded area represents the theoretical force if $T_\text{evap}$ is up to 1~K higher than the mode temperature. Inset: Displacement spectrum of the flexural mode at $0.7~\rm K$ and $2~\rm K$ with a coherent drive applied via optical amplitude modulation. The response to coherent drive is shown to increase with the presence of superfluid helium.
}
\end{figure}

%%%%%%%%%%%%%%%%%%%%%%%%%%%%%%%%%%%%%%%%%%%%%%%%%%%%%%%%%%%%%%%%%%%%%%%%%%%%%%%%%%%%%%%%%%%%%%%%%%%%%%%%%%
%Boil-off
%%%%%%%%%%%%%%%%%%%%%%%%%%%%%%%%%%%%%%%%%%%%%%%%%%%%%%%%%%%%%%%%%%%%%%%%%%%%%%%%%%%%%%%%%%%%%%%%%%%%%%%%%%

%With increasing laser power thermal effects eventually begin to degrade the system. In the case of our superfluid based system this thermal degradation corresponds to boil-off of the superfluid film. To measure this effect we monitor the mode temperature as the laser power is varied by over two orders of magnitude (Fig.~\ref{Fig2}(b)).
To investigate the effects of optical absorption on the temperature of the microtoroid, we determined the integrated power spectral density of the mechanical mode as a function of laser power. The temperature was found to increase with increasing laser power, eventually causing a boil-off of the superfluid film, as shown in Fig.~\ref{Fig2}(b). 
%The mode temperature increases from 510~mK to 730~mK as the laser power is increased over two orders of magnitude from 10~nW to 2.1~$\mu$W. 
As the laser power is increased over two orders of magnitude from 10~nW to 2.1~$\mu$W the mode temperature increases only modestly from 510~mK to 730~mK. 
Above 2.2~$\mu$W the mode temperature jumps sharply to 3~K, indicated by the red shaded region in Fig.~\ref{Fig2}(b).
This threshold behaviour manifests due to Landau's critical velocity, which sets an upper limit on the superfluid flow rate (see Supplementary Information). This results in a thermal run-away process, wherein the superfluid can no longer be replenished at the periphery of the microtoroid as fast as it evaporates and therefore boils off completely. The microtoroid is then no longer effectively thermally anchored to the cryostat, and the final mode temperature is dominated by laser heating.
%This large temperature increase is a manifestation of the limitation of flow through the pedestal by the critical velocity of superfluid. The resulting effect is thermal run-away since the superfluid cannot be replenished at the periphery fast enough for thermalization. 
%The microtoroid is then no longer thermally anchored to the cryostat with a final temperature dominated by laser heating.

%%%%%%%%%%%%%%%%%%%%%%%%%%%%%%%%%%%%%%%%%%%%%%%%%%%%%%%%%%%%%%%%%%%%%%%%%%%%%%%%%%%%%%%%%%%%%%%%%%%%%%%%%%
%Superfluid Enhanced Optical Forces
%%%%%%%%%%%%%%%%%%%%%%%%%%%%%%%%%%%%%%%%%%%%%%%%%%%%%%%%%%%%%%%%%%%%%%%%%%%%%%%%%%%%%%%%%%%%%%%%%%%%%%%%%%

%To investigate optical forces applied to the flexural mode a network analysis was performed, where the drive is applied via amplitude modulating the injected field (see Fig.~\ref{ExpSetup}(a)), as the cryostat temperature was varied.
To investigate the optical forces present in the system a constant optical amplitude modulation was applied at the frequency of the flexural mode as the cryostat temperature was varied (Fig.~\ref{Fig3}). 
%This was done using a network analyser which amplitude modulates the injected field, hence driving the microtoroid motion (see Fig.~\ref{Fig1}(a)).
This applies resonant forces on the mode, both through radiation pressure, and, below the superfluid transition temperature, superfluid flow.
The mechanical response to this drive was measured via homodyne detection of the phase quadrature of the output field.
%Details
%Drive toroid motion and measure the response via homodyne detection of phase of output field.
At temperatures above the superfluid transition the optical force originates from radiation pressure alone and is essentially independent of temperature (see right inset to Fig.~\ref{Fig3}). However, upon formation of a superfluid layer, indicated by the blue shaded region in Fig.~\ref{Fig3}, the response of the flexural mode to the laser drive abruptly increases by 21~dB to a maximum of 540~fN (left inset to Fig.~\ref{Fig3}). Taking into account the poor overlap between the flexural mode and the radial evaporative force (0.037\% calculated from FEM modelling) gives a total superfluid photoconvective force of 1.46~nN.
This increased response in the presence of superfluid is in good agreement with theoretical predictions, corresponding to a superfluid photoconvective force that is a factor of eleven larger than radiation pressure. 
%As described earlier, this increased response is due to superfluid evaporation resulting in a recoil force that is applied to the mechanical mode and is in good agreement with theory (see Supplementary Information). 
%
%Using the relative strengths of the evaporative recoil and radiation pressure forces (see Supplementary Information) a theoretical fit to the combined forcing can be obtained. 
%With the absolute value of the radiation pressure drive as the only free parameter, Fig.~\ref{Fig3} (solid line) shows good agreement between theory and experimental results. 
%The downward trend of applied forcing below the transition temperature (seen in Fig.~\ref{Fig3}) arises from the 
The measured superfluid convective force decreases in magnitude as the temperature is reduced away from the transition temperature (see Fig.~\ref{Fig3}). This occurs because of the reduced RMS velocity of the evaporated atoms (see Eq.~\ref{v_rms}), with colder atoms contributing less recoil to the microtoroid.
This behaviour is accurately predicted by our model, where the superfluid evaporation temperature in Eq.~\ref{evapForce} is equated to the measured microtoroid mode temperature $T_m$. However, the observed superfluid forces are found consistently to be larger than predicted by the model, with a maximum deviation of approximately 60\%. We attribute this discrepancy to a temperature differential existing between the evaporated atoms and the helium film.
It has been shown that helium atoms evaporated from a superfluid thin film have a temperature that is up to 1~K hotter than the film temperature, dependent on the total heat applied to the liquid~\cite{Hyman_PhysRev69, Andres_PhysRevA73}. To account for this phenomenon we have included in Fig.~\ref{Fig3} a theoretical band (pink shading) showing the expected applied force for atoms that are evaporated with temperatures $T_\text{evap}$ ranging from the mode temperature $T_{\text{m}}$ to $T_\text{m}+1$~K.

\begin{figure}[t]
\begin{center}
\includegraphics[width=0.95\columnwidth]{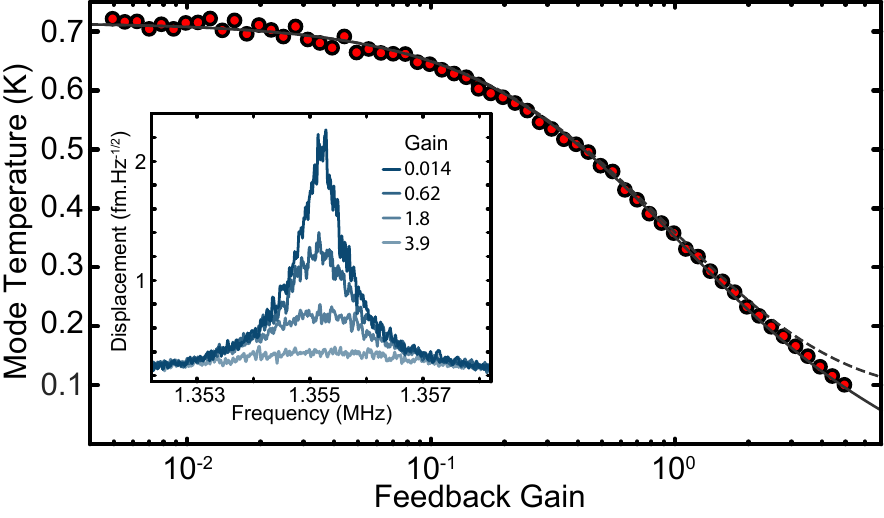}
\end{center}
\caption{\label{Fig4} %(a) Mode temperature of the flexural mode as the probe laser is increased from $10~\rm nW$ to $3.3~\rm \mu W$. Below $2.2~\rm \mu W$ the temperature increases slightly as the laser power is increased. Above $2.2~\rm \mu W$ the superfluid boils off causing a sharp rise in mode temperature. (b) 
Feedback cooling of the flexural mode from $715~\rm mK$ to $137~\rm mK$ using superfluid mediated photothermal forcing. With a fixed probe power of $1.9~\rm\mu W$ the feedback gain is varied over three orders of magnitude, showing good agreement with the estimated mode temperature from in-loop measurements (solid line). The out-of-loop mode temperature is then inferred by a transformation (dashed line) that is derived in the Supplementary Information. Inset: Displacement spectrum of the flexural mode with varying feedback gain.
}%44\% increase to 2uW %photocurrent is filtered, amplified and applied as an optical amplitude modulation,
\end{figure}

As a specific example of an application that takes advantage of the enhanced optical force provided by the superfluid, we perform feedback cooling on the microtoroid mode. This is done by passing the homodyne photocurrent through various filter and amplification stages, then feeding it into an amplitude modulator placed before the microtoroid (see \fig{Fig1}(a)). Provided the phase of the feedback loop has been chosen correctly to provide a force that opposes the velocity of the mode, the thermal motion of the 1.35~MHz flexural mode is reduced via cold damping~\cite{Cohadon_PRL99}. Figure \ref{Fig4} shows that as the feedback gain is increased the microtoroid mode temperature decreases in excellent agreement with theory~\cite{Lee_PRL10}. The flexural mode is thus cooled from 715~mK to 137~mK, with a final occupancy of $n=2110 \pm 40$ phonons; constrained primarily by the optomechanical coupling rate to the flexural mode. %The maximum cooling achieved here is limited by low cooperativity, which is constrained by the weak optomechanical coupling rate of the flexural mode and the amount of laser power that can be applied.

While we already demonstrated an order of magnitude improvement in optical force over radiation pressure using evaporative recoil in thin superfluid films, for completeness we discuss in the supplementary information the magnitude of the forces arising from superflow and normal fluid counterflow during heat transport in bulk superfluid (as illustrated in Fig. \ref{fig4} (a)). By choosing the right experimental parameters, we show that the force could be further increased by over an order of magnitude compared to the thin film case, enabling optical forces more than 2 orders of magnitude larger than what is achievable with radiation pressure even in a high finesse cavity. 
 
%. This has the effect of increasing the number of atoms ($N$) required to remove the thermal energy from photon absorption. 
%Normal fluid counter flow velocities as low as xxx have been achieved [cite], yyy orders of magnitude smaller than the RMS thermal velocity at 1 K. blah blah…  This would allow a zzz increase in efficiency...
% 
%Even with these improvements, in practice, the total force will be constrained beneath unity by the conversion efficiency from photon energy to superfluid flow kinetic energy, and also by technical issues such as alignment of the atomic stream to the mechanical element. 
%However, the superfluid fountain effect has been shown to convert heat to flow with an efficiency as high as $10\%$~\cite{Hofmann_Patent87}. 
%usually limited by the quality of the superleak that prevents normal fluid counterflow. 
%Using superfluid helium as the atomic stream the mass equivalence required for high efficiency from Eq.XXX is easily attained, for example a femtogram optomechanical element would require a volume of superfluid helium less than $<0.01\rm mm^3$.

\begin{figure}[t]
\begin{center}
\includegraphics[width=0.95\columnwidth]{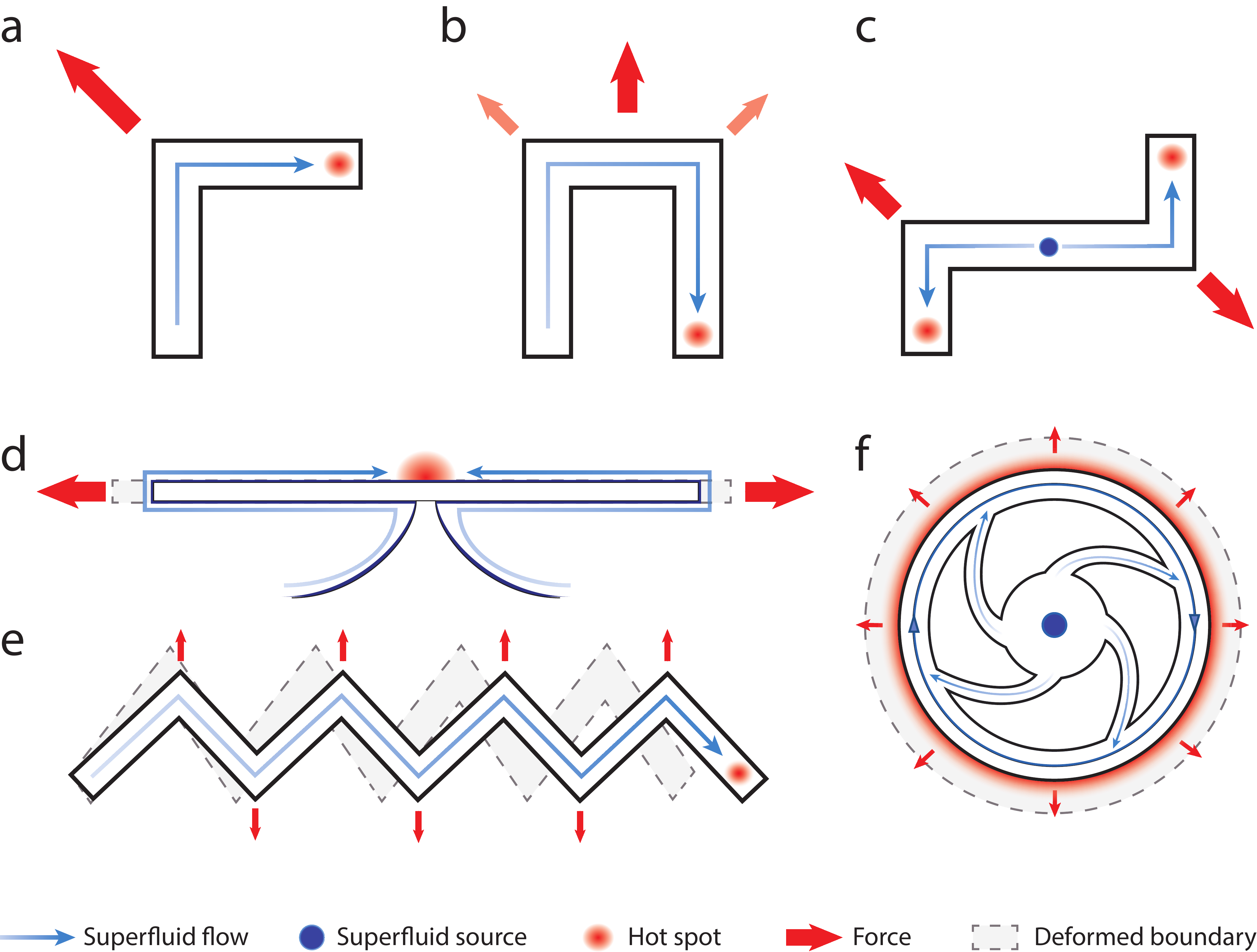}
\end{center}
\caption{\label{SFforces} By varying the geometry, superfluid flow could be used to apply a large range of different forces including: (a) \& (b) linear force, (c) torque, (d--f) examples of geometries designed to efficiently leverage forces from superfluid flow. (d) Expansive force that is efficiently coupled to a microdisk/microtoroid, (e) compressive force, and (f) long lived centrifugal forces resulting from persistent current flow.
}
\end{figure}

%%%%%%%%%%%%%%%%%%%%%%%%%%%%%%%%%%%%%%%%%%%%%%%%%%%%%%%%%%%%%%%%%%%%%%%%%%%%%%%%%%%%%%%
%Context (COMS and )
%%%%%%%%%%%%%%%%%%%%%%%%%%%%%%%%%%%%%%%%%%%%%%%%%%%%%%%%%%%%%%%%%%%%%%%%%%%%%%%%%%%%%%%
%Efficient feedback (COMS)
The strongest known optical actuation capabilities are provided by the photothermal interaction~\cite{Restrepo_CRP11}. Indeed, the first optomechanical system to demonstrate cooling via dynamical backaction was based on this mechanism~\cite{Metzger_Nat04}. Further, it has been shown that photothermal forces should enable optomechanical cooling to the ground state, without requiring sideband resolution~\cite{Restrepo_CRP11}; thus enabling efficient cooling of low frequency mechanical oscillators. However, accessing large photothermal forces requires strong optical absorption and a large thermal expansion coefficient, limiting devices to specific materials and geometries, and additionally precludes cryogenic operation as the thermal expansion coefficient of most materials reduces by several orders of magnitude when cooled to cryogenic temperatures. Furthermore, the characteristic bandwidth is defined by the typically slow thermalization rate of the device material. Superfluid convective and evaporative forces could alleviate these constraints, allowing new regimes to be realized characterized by fast, strong actuation; with the potential to incorporate superfluid-enhanced optical forcing into existing cryogenic optomechanical systems.

Superfluid convective and evaporative forces may find applications where strong optical forces are required in the absence of an optical cavity; for example in photonic circuits~\cite{Li_Nat08, Li_NatPhot09, Roels_NatNano09}, or cryogenic MEMS~\cite{Waldis_QuantElec07}. It should also be possible to design systems where the heat source is applied at a location spatially remote from the resultant force, and in a range of geometries to apply not only forces, but also torques, at the microscale (see Fig.~\ref{SFforces}). This could be advantageous in applications where the device is highly-sensitive to temperature fluctuations.

%%%%%%%%%%%%%%%%%%%%%%%%%%%%%%%%%%%%%%%%%%%%%%%%%%%%%%%%%%%%%%%%%%%%%%%%%%%%%%%%%%%%%%%
%Conclusion
%%%%%%%%%%%%%%%%%%%%%%%%%%%%%%%%%%%%%%%%%%%%%%%%%%%%%%%%%%%%%%%%%%%%%%%%%%%%%%%%%%%%%%%
In conclusion, we have demonstrated photoconvective forcing of a mechanical oscillator based on superfluid flow and recoil. This enables large optical forces at cryogenic temperatures, in contrast to photothermal forcing mechanisms that are generally precluded from cryogenic operation.
%a novel type of optical force based on the recoil of evaporated superfluid helium-4, enabling large forces to be accessible for microscopic systems at cryogenic temperatures. 
Furthermore, the exceptionally high thermal conductivity of superfluid helium provides good thermal anchoring to the environment while permitting fast optical forces to be realized. The self assembling nature of superfluid helium means this technique may be relatively straightforwardly incorporated into other cryogenic optomechanical systems.

{\it Acknowledgments}: This research was funded by the Australian Research Council Centre of Excellence CE110001013. WPB is supported by the Australian Research Council Future Fellowship FT140100650. Device fabrication was undertaken within the Queensland Node of the Australian Nanofabrication Facility.

\end{document}